\documentclass[twocolumn,showpacs,preprintnumbers,amsmath,amssymb]{revtex4}

\usepackage{graphicx}
\usepackage{dcolumn}
\usepackage{bm}
\usepackage{color}

\begin{document}


\title{Origin of heavy tail statistics in equations of the Nonliner Schr\"odinger type:  \\ an exact result}

\author{M. Onorato$^{1,2}$, D. Proment$^3$, G. El$^4$, S. Randoux$^5$, P. Suret$^5$,}
\affiliation{$^1$  Dipartimento di Fisica, Universit\`a degli Studi di Torino, 10125 Torino, Italy}
\affiliation{$^2$  Istituto Nazionale di Fisica Nucleare, INFN, Sezione di Torino, 10125 Torino, Italy}
\affiliation{$^3$  School of Mathematics, University of East Anglia, Norwich Research Park, Norwich, NR4 7TJ, United Kingdom}
\affiliation{$^4$ Department of Mathematical Sciences,
Loughborough University, Loughborough
Leicestershire, LE11 3TU, United Kingdom}
\affiliation{$^5$ Laboratoire de Physique des Lasers, Atomes et Molecules, Universit\'e de Lille, UMR-CNRS 8523, France}

\date{\today}

\begin{abstract}
We study the formation of extreme events in incoherent systems described by  envelope equations, such as the Nonliner Schr\"odinger equation.
 We derive an identity that 
relates the evolution of the kurtosis (a measure of the relevance of the tails in a probability density function) of the wave amplitude to the rate of change of the width of the Fourier spectrum of the wave field. 
The result is exact for all dispersive systems characterized by a nonlinear term of the 
form of the one contained in the Nonlinear Schr\"odinger equation. Numerical simulations are also performed to confirm our findings. 
Our work sheds some light on the origin of 
rogue waves in incoherent dispersive nonlinear media ruled by local cubic nonlinearity.
\end{abstract}

\maketitle
Processes that lead to the formation of heavy  tails \cite{resnick2007heavy,beirlant2006statistics,malamud2004tails} in the Probability Density Function (PDF)
are of wide interest in many physical contexts \cite{kharif2009rogue,onoratoetal04,barabasi2005origin,solli2007optical,walczak2015optical,toenger2015emergent,pierangeli2015spatial}. It is well known that in homogeneous conditions, if the central limit theorem applies,  a wave  linear  dispersive system characterized by a large number of incoherent waves is described by a Gaussian statistics; in the latter situation 
extreme events can still appear but they are very rare, and their probability of appearance can be derived exactly, \cite{rice1945mathematical,LH52}.

In the field of ocean waves and nonlinear optics, it has been established that 
the presence of nonlinearity on top of dispersion can lead to changes in
 the statistical properties of the system. Often rogue waves and the associated large tails in the PDF can be observed in experiments or numerical simulations from an initially incoherent wave field \cite{onorato2013rogue,dudley2014instabilities};  the situations characterized by  lower probability of  extreme events than the Gaussian predictions can also be encountered
 \cite{janssen03,Randoux:14}. In all those cases the nonlinearity plays a key role in 
 creating correlations among modes that ultimates in a deviations from Gaussian statistics. Given a partial differential equation for a dispersive nonlinear system and given an initial condition characterized by some statistical properties,  it would be desirable to establish  the PDF of the wave field at different times. This is not an easy task, and nowadays, some results can be achieved only by making a number of assumptions, as it is done in the wave turbulence theory \cite{nazarenko2011wave}. 
 
In this Letter we present a very simple identity which can be derived  from a family of universal nonlinear dispersive partial differential equations that allows one  to relate the changes in the statistical properties of the wave field 
to the changes of its Fourier spectrum.  
Specifically, our focus is  on the kurtosis, i.e. the fourth-order moment of the PDF  which measures the relevance of the tails of the distribution with respect to the core. Large values of kurtosis  imply the  presence of heavy tails in the distribution and higher probability of extreme events. 
We show analytically, without any approximation, that an increase of the spectral bandwidth (typical of a  
nonlinear evolution) results in an increase/decrease of extreme events in focusing/defocusing regime.
Here,  we will first discuss the 1D+1 integrable Nonlinear Schr\"odinger (NLS) equation problem  and then we will extend the result to non-integrable NLS type of equation in 2D+1 and confirm our results with extensive 
numerical simulations.

{\it One-dimensional case.} 
The NLS equation is a universal model for describing nonlinear dispersive waves. For the present discussion, we will consider the NLS equation written as follows:
\begin{equation}
i \frac{\partial A}{\partial x}= \beta\frac{\partial^2 A}{\partial t^2}+\alpha |A|^2A, \label{NLS}
\end{equation}
where $\alpha$ and $\beta$ are two constant coefficients that  depend on the physical problem considered. 
If $\alpha \beta > 0$ then the equation is known to be of focusing type, while if  $\alpha \beta < 0$ 
the equation is defocusing. Note that equation (\ref{NLS}) is written as an evolution equation in space rather than in time;
this notation is common in nonlinear optics and it is also suitable in hydrodynamics for describing the evolution of 
waves in wave tank experiments. The general problem that one wish to answer is the following: given an incoherent time series characterized by some statistical properties at one boundary of the domain, what is the PDF of the intensity of the wave field along the tank or along the fiber? Will rogue waves appear?
We stress that our goal  here is not to establish  the validity of the NLS equation
in a specific field but to highlight a fundamental mechanism that leads to the formation of extreme or rogue waves. 

We start by the definition of the kurtosis:
\begin{equation}
\kappa(x,t)=  \frac{\langle |A(x,t)|^4\rangle}
{\langle |A(x,t)|^2\rangle^2}, \label{def_kurt}
\end{equation}
where $\langle...\rangle$ denotes the ensemble average over different realisation of the same process characterized by different phases and  Fourier amplitudes. The kurtosis is the fourth order moment of the PDF and it weights the relevance of its tails: heavy tails implies the presence of extreme/rogue waves.
 In the recent years, some interesting approaches have been developed in order  to estimate the kurtosis in  NLS type of equations
\cite{janssen03,nazarenko2011wave,janssen2009some,annenkov2006role,fedele_kurtosis_2015}; up to our knowledge, they  all rely on a closure.
Being a nonlinear problem with cubic nonlinearity, the evolution of the fourth-order moment is related to the evolution of the sixth-order moment and so 
on, and a quasi-Gaussian approximation is assumed to get a closed form for the kurtosis. Such a closure is, of course, questionable in the presence of strong nonlinearity.  
The theoretical development that we 
present in this Letter is different: it does not require a closure and it is valid for an arbitrary nonlinearity.
It is based on the (now trivial) intuition that the nonlinear 
part of the Hamiltonian is strictly related to the kurtosis. 
Indeed, the following Hamiltonian density is conserved for equation (\ref{NLS}) at every $x$:
\begin{equation}\label{eq:3}
H=\frac{1}{T}\int_0^T
\beta\left|\frac{\partial A}{\partial t}\right|^2dt-
\frac{1}{T}\int_0^T \frac{\alpha}{2}\left|A\right|^4dt.
\end{equation}
We now take the ensemble average of the above equation to get:
\begin{equation}\label{eq:4}
\langle H\rangle=\frac{1}{T}\int_0^T
\beta\langle\left|\frac{\partial A}{\partial t}\right|^2\rangle dt-
\frac{1}{T}\int_0^T \frac{\alpha}{2}\langle \left|A\right|^4 \rangle dt.
\end{equation}
Assuming that the $A(x,t)$ is a statistically stationary process in the interval  $[0,T]$,  
then  
$\langle \left|A\right|^4 \rangle$ is time independent and 
Eq. (\ref{eq:4})  can be re-written as follows:
\begin{equation}
\frac{\langle H\rangle}{\langle N\rangle}=\beta  \Omega(x)^2-
\frac{\alpha}{2}\langle N \rangle \kappa(x) \label{hamilt_kurt}
\end{equation}
with $\langle N\rangle$ being the ensemble average of the number density of particles defined as
\begin{equation}
\langle N\rangle=\frac{1}{T}\int_0^{T}{\langle|A(x,t)|^2\rangle}dt=\langle|A(x,t)|^2\rangle,
\end{equation}
(the last equality holds for a statistical stationary process)
and
\begin{equation}\label{eq:7}
\Omega(x)=\sqrt{\frac{ \sum_{n} \langle (\frac{2 \pi}{T}n)^2|A_n(x)|^2\rangle}{\sum_n  \langle |A_n(x)|^2\rangle}},
\end{equation}
with
$A_n(x)$ being the Fourier coefficients defined as
\begin{equation}
A_n(x)=\frac{1}{T}\int_0^T A(x,t) e^{-i \frac{2\pi}{T} n t }dt.
\end{equation}
Note that periodic boundary conditions in $t$ have been assumed.  The quantity
 $\Omega(x)$ defined by Eq. (\ref{eq:7}) is nothing but the definition of the spectral bandwidth.
Evaluating the expression (\ref{hamilt_kurt}) at $x=x_0$ (initial condition) and at a generic point $x$, after eliminating $\langle H\rangle/\langle N\rangle$ from the two resulting equations, we get
the following exact relation (note that $\langle H \rangle $ and $\langle N \rangle$ do not depend on space and time):
\begin{equation}
\kappa(x)=\kappa(x_0)+2 \frac{\beta}{\alpha} \frac{1}{ \langle N \rangle} \left[ \Omega(x)^2-\Omega(x_0)^2 \right]. \label{kurtosis}
\end{equation}
%
Equation (\ref{kurtosis}) implies that the variation of the kurtosis is
directly related to the variation of the spectral bandwidth. For example, in deep water gravity waves and in the anomalous dispersion regime in optics $\beta/\alpha>0$ 
 the increase of the spectrum is accompanied by an increase of the kurtosis and, as it will be seen below, 
 to the formation  of large tails in the spectrum.  
 On the contrary, when the water depth, $h$, decreases below a critical 
value ($k_0 h<1.36$, with $k_0$ the wave number of the carrier wave) or when the fiber is characterized by a regular dispersion, then the ratio $\beta/\alpha$ becomes negative (the NLS is of defocusing type) and an increase of the spectral 
 bandwidth leads to a decrease of the kurtosis (less extreme events than predicted by the Gaussian distribution).
 If the spectrum is frozen, the kurtosis does not evolve. 
 
\begin{figure}
\centerline{\includegraphics[width=7cm]{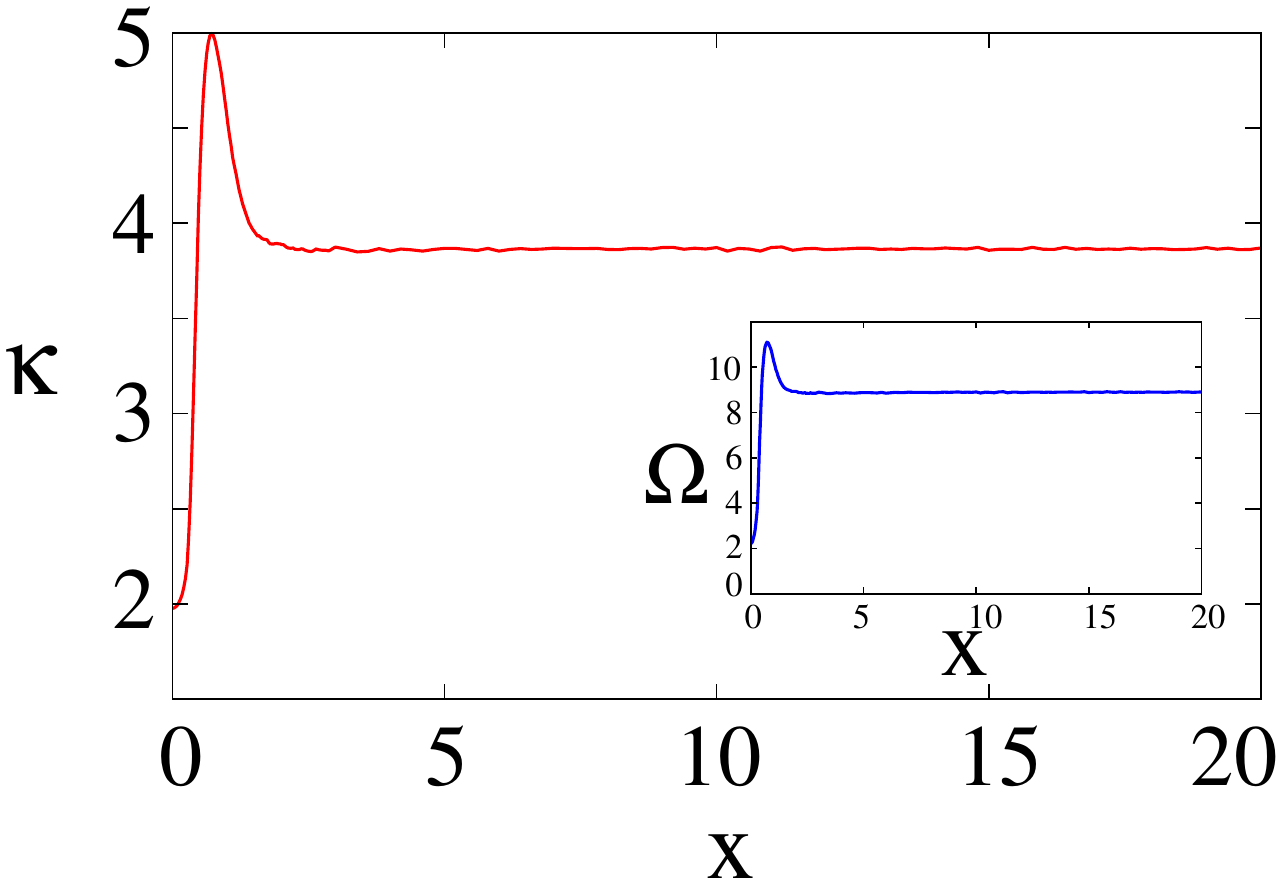}}
 \caption{Evolution of  the kurtosis  for the focusing NLS equation ($\alpha=2$, $\beta=1$). In the inset the evolution of  the spectral bandwidth is shown.}
 \label{fig_kurt_nls_foc}
 \end{figure}

\begin{figure}
\centerline{\includegraphics[width=7cm]{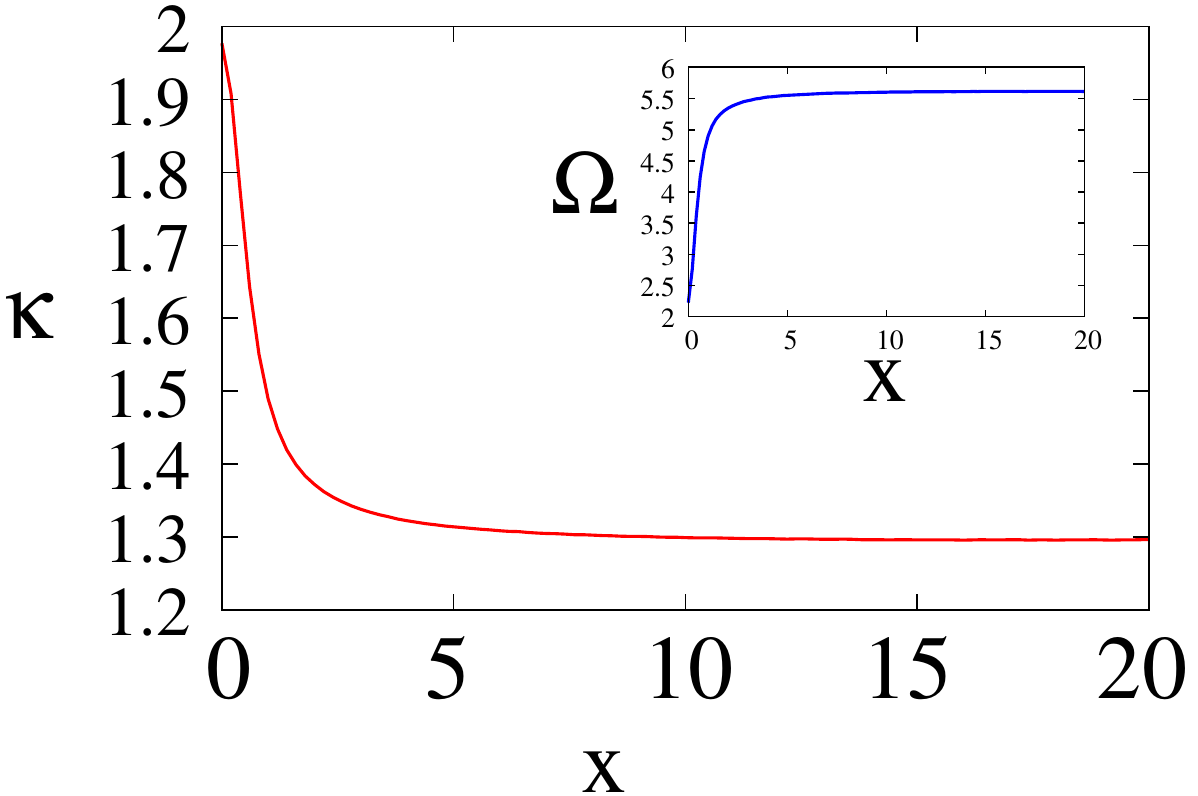}}
 \caption{Evolution of the kurtosis for the defocusing NLS equation. ($\alpha=-1$, $\beta=1$). In the inset the evolution of  the spectral bandwidth is shown.}
 \label{fig_kurt_nls_defoc}
 \end{figure}

 In what follows, we consider a few numerical examples; without loss of generality, we solve the NLS equation  (\ref{NLS}) with $\alpha=\pm 2$ and $\beta=1$, starting from an 
 initial condition characterized by the following  frequency Fourier spectrum:
 \begin{equation}
 A_n(x=0)=\sqrt{a_0e^{-\frac{4\pi^2 n^2}{T^2\sigma^2}}}e^{i \phi_n},
 \end{equation}
where the phases $\phi_n$ are distributed uniformly in the $[0,2\pi)$ interval. 
The numerical simulations are performed by using a pseudo-spectral
method with a step-adaptative algorithm permitting to reach a specified 
level of numerical accuracy; a box of size $T=257.36$ has been discretized in $4096$ 
points. The numerical values of $a_0$ and $\sigma^2$ are $1.129$  and $10^4$, respectively. 
The statistical properties of the random wave fields are computed 
from an ensemble of $10^4$ realisations of the random initial condition. 
Because of the latter choice, the PDF of the real and imaginary part of $A(t,x=0)$ is Gaussian, the PDF of $|A(t,x=0)|$ is  distributed according to  the Rayleigh distribution having kurtosis equal to 2, and the PDF of the intensity $|A(t,x=0)|^2$ is exponential. 
In Figures  \ref{fig_kurt_nls_foc} and 
 \ref{fig_kurt_nls_defoc} we show the kurtosis and the spectral bandwidth as a function of the evolution variable $x$ for the focusing, $\alpha=2$, and the defocusing case, $\alpha=-2$, respectively. 
\begin{figure}
\centerline{\includegraphics[width=8cm]{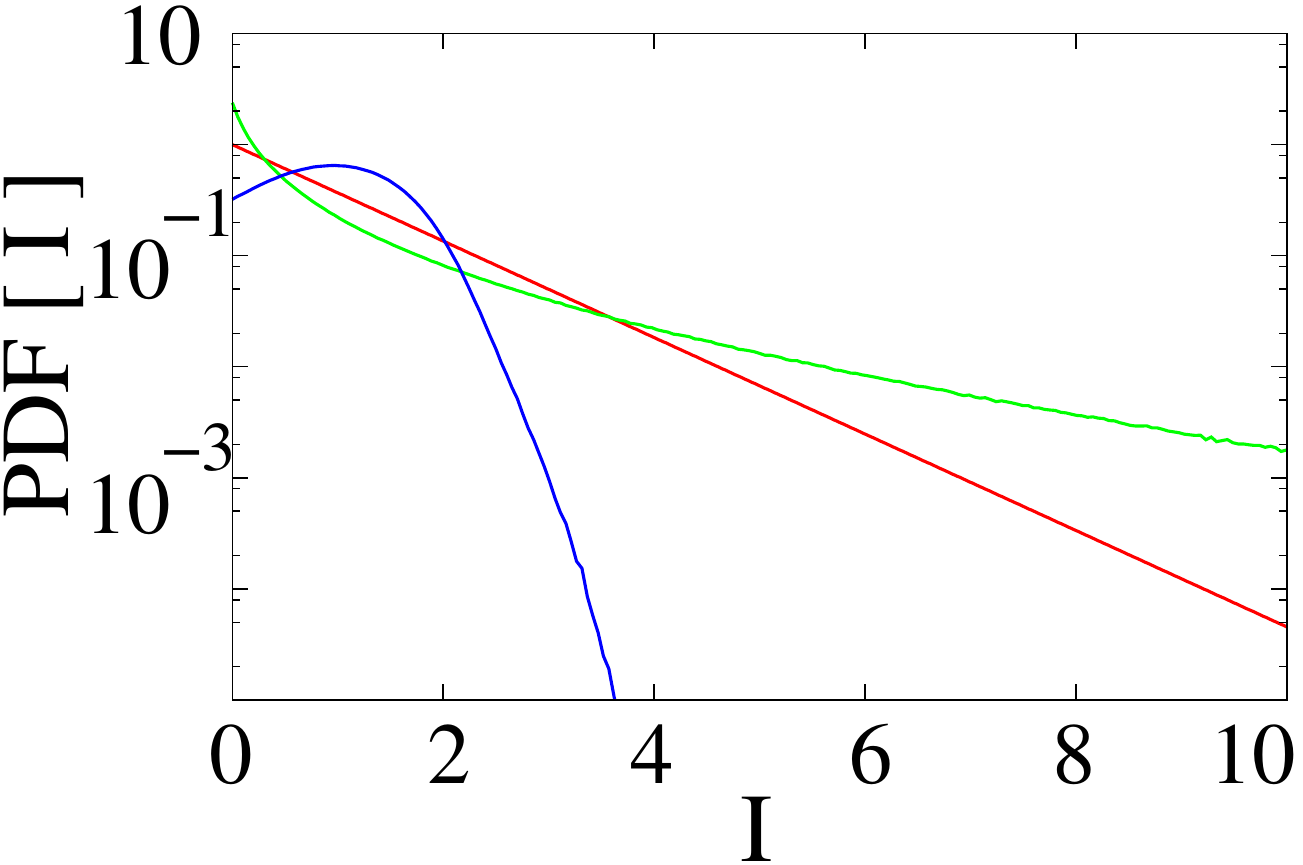}}
 \caption{PDF of  the normalized intensity $I=|A(t,x)|^2/N$ for both focusing (green line) and defocusing (blue line) NLS equation calculated for $x>20$. The exponential distribution is also shown as a red line.}
 \label{fig_PDF_nls}
\end{figure}
It is interesting to note that, regardless of the sign of $\alpha$, the spectral bandwidth  always increases; however, while 
the kurtosis increases in the focusing case, it decreases in the defocusing one. As was mentioned, high values of kurtosis, implies heavy tails in the PDF. Indeed, in Figure \ref{fig_PDF_nls} the PDF of the normalized intensity $I=|A(t,x)|^2/N$ computed  
after the kurtosis has reached 
an equilibrium state,  
 ($x>20$), 
is shown for both the focusing and the defocusing case. Numerical results are compared with the exponential distribution $e^{-I}$, the expected distribution for a system of linear incoherent waves. 
Deviations from such distribution are observed for both cases, however, consistently with our derivation, the focusing case shows  heavy tails, while the defocusing one, the distribution is below the exponential prediction.
%
 %
 %
 
{\it Two-dimensional case.}
We now consider the NLS in two horizontal dimensions written as an evolution equation in time:
\begin{equation}
i \frac{\partial A}{\partial t}= \left(\beta \frac{\partial^2 A}{\partial x^2}+
 \gamma \frac{\partial^2 A}{\partial y^2}\right)+\alpha |A|^2A \label{NLS_2D}
\end{equation}
with $\gamma=\pm 1$.
In the water wave context, equation  (\ref{NLS_2D}) with  $\alpha=\beta=1$ and $\gamma=-1$ arises in the deep water regime and it describes the evolution of the complex wave envelope in  weakly nonlinear and narrow band (both in the direction of propagation and in the 
its transverse direction) approximations. 
The second derivative in the  $y$ direction plays the role of diffraction and the equation is known as the Hyperbolic NLS.
On the other hand,  equation (\ref{NLS_2D}) with the choice of $\gamma=\beta=-1$ and $\alpha=1$, also known as the  defocusing Gross-Pitaevskii equation (GPE), describes for instance the dynamics of a two-dimensional Bose-Einstein condensate.
%
%
We now  assume  that the system in homogeneous in the domain $L_x\times L_y$ and we follow the same procedure as in the one dimensional case.
Keeping in mind that now the 
kurtosis evolves in time and it is defined as an average over space (rather than time), the same reasoning as before  can be applied to get:
\begin{equation}
\begin{split}
&\kappa(t)=\kappa(t_0)+ \frac{2}{\alpha \langle N \rangle}\times\\
& \left\{ \beta \left[K_x(t)^2-K_x(t_0)^2\right]+\gamma \left[K_y(t)^2-K_y(t_0)^2)\right] \right\}, \label{kurtosis_2D}
\end{split}
\end{equation}
where
\begin{equation}
K_x(t)=\sqrt{ \frac{ \sum_{k,l} \langle \left(\frac{2 \pi}{L_x} k\right)^2|A_{k,l}(t)|^2\rangle}{\sum_{k,l} \langle |A_{k,l}(t)|^2\rangle}},
\end{equation}
and $K_y(t)$ is defined in a similar fashion with the only difference that in the brackets  $L_x$ is replaced by $L_y$
 and $k$ with $l$.
  
 As in the one dimensional case, we show  some instructive numerical simulations of equation 
 (\ref{NLS_2D}). For all cases considered, the initial condition is characterized by the 
following Fourier spectrum:
\begin{equation}
A_{k,l}(t=0)=\sqrt{a_0e^{-\left(\frac{2\pi}{L}\right)^2\frac{k^2+ l^2}{\sigma}}}e^{i \phi_{k,l}}
\label{initial_2D}
\end{equation}
with $a_0=7.8\times10^{-5}$, $\sigma=\sqrt{10}$, $L=L_x=L_y=512$
and phases are taken as randomly distributed. 
Numerical simulations are performed 
with a resolution of $1024\times1024$ with $\Delta x=\Delta y=0.5$. To improve the statistical convergences, 10 different simulations are performed for each case with different initial random phases.

\begin{figure}
\centerline{\includegraphics[width=8cm]{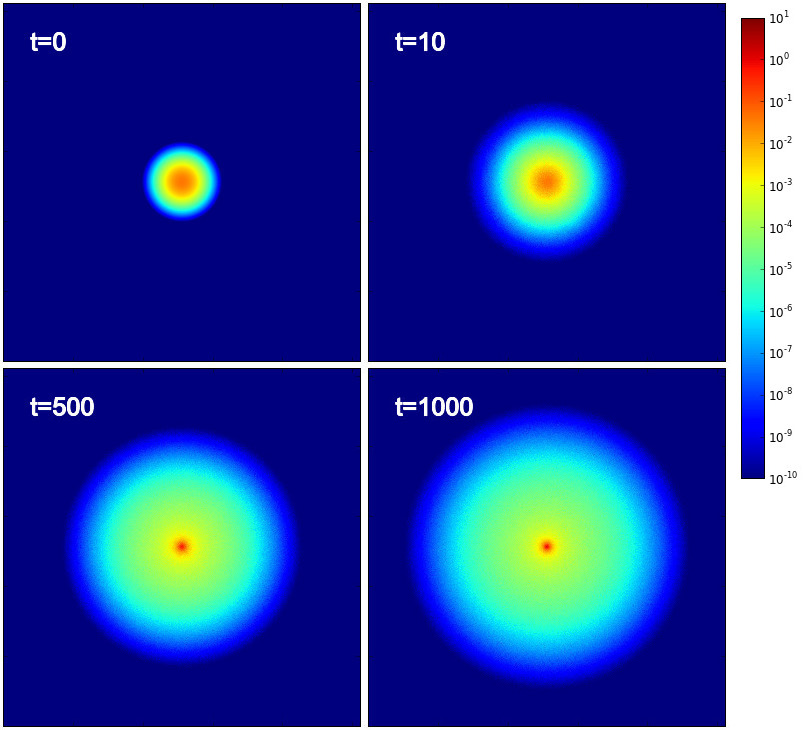}}
 \caption{Evolution of the spectrum for the GPE at different times. The initial conditions are provided in equation (\ref{initial_2D})
  }
 \label{fig_spectra_GPE}
 \end{figure}

\begin{figure}
\centerline{\includegraphics[width=8cm]{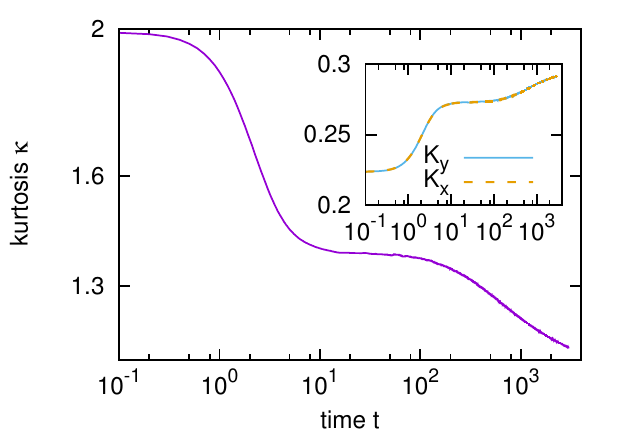}}
 \caption{$\kappa(t)$  as a function of time for numerical simulations of the 
 GPE. The initial conditions are provided in equation 
 (\ref{initial_2D}). In the inset the evolution of the spectral bandwidth is shown.
 }
 \label{fig_kurt_GPE}
 \end{figure}

We start by considering the GPE: generally, given an initial condition localized in Fourier space, the tendency is to observe a broadening of the spectrum, thus, due to the fact that $\beta =\gamma =-1$, according to 
equation (\ref{kurtosis_2D}) we expect to observe a decrease of the kurtosis.
Indeed, in Figure \ref{fig_spectra_GPE}, we show a density plot of the two dimensional 
Fourier spectrum at $t=0,\;10,\; 500,\; 1000$. It is interesting to observe that the spectrum broadens isotropically and a condensate at the mode $ (k,l)=(0,0)$ forms at large times (red spot in the Figure \ref{fig_spectra_GPE}), see \cite{nazarenko2014bose,connaughton2005condensation} for details. 
The initial kurtosis, shown in  Figure \ref{fig_kurt_GPE}, decreases from the value of 2:  starting from a spectrum characterized by random phases,  extreme events are not expected in the defocusing GPE.
\begin{figure}
\centerline{\includegraphics[width=8cm]{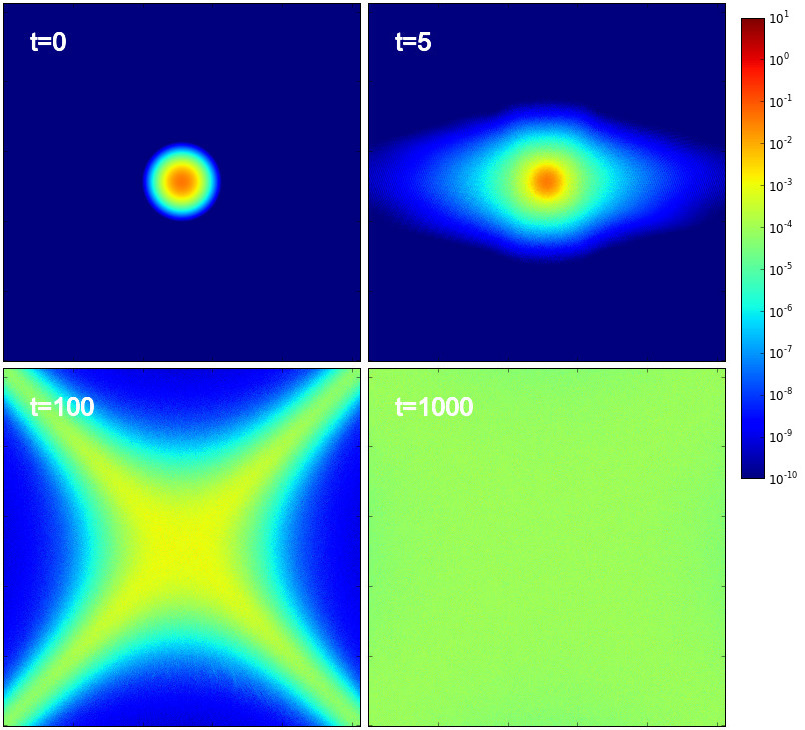}}
 \caption{Evolution of the spectrum for the hyperbolic NLS equation at different times. The initial conditions are provided in equation (\ref{initial_2D})
  }
 \label{fig_spectra_hyperbolic}
 \end{figure}
 \begin{figure}
\centerline{\includegraphics[width=8cm]{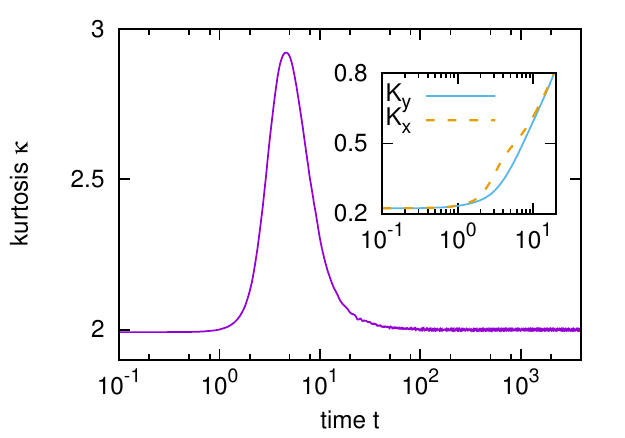}}
 \caption{$\kappa(t)$  as a function of time for numerical simulations of the 
 GPE. The initial conditions are provided in equation 
 (\ref{initial_2D}). In the inset, the evolution of the spectral bandwidth is shown.
 }
\label{fig_kurt_hyperbolic}
\end{figure}
The situation is different for the hyperbolic NLS where the equation is focusing in $x$ and defocusing in the $y$ direction.
Because of the opposite signs in the linear terms, we expect an initial 
non-isotropic evolution. Indeed, as shown in Figure \ref{fig_spectra_hyperbolic},
the spectrum evolves more rapidly in the $k_x$ direction, probably due to 
some fast evolution related to an instability of the modulational instability type, see \cite{alber78}. This results in a fast increase of the kurtosis, see Figure \ref{fig_kurt_hyperbolic} up to $t=5$. After this initial transient, the spectrum grows also in the transverse direction and the value of the kurtosis reduces accordingly. The signs of $\beta$ and $\gamma$  in equation (\ref{kurtosis_2D}) are now opposite, therefore they have a competing effect: an increase of the spectral bandwidth in the $k_x$ direction leads to 
a increase of the kurtosis, while the opposite happens for an increase of the spectral bandwidth in the $k_y$ direction. After some times the spectrum appears again isotropic  and the contribution in (\ref{kurtosis_2D}) from the spectral bandwidth cancels out and the kurtosis tends to its Gaussian value. 
A snapshot of the intensity of the wave field taken at the time when the kurtosis has a maximum 
 is reported in Fig. \ref{fig_rogue}: clearly, the field is characterized by the presence of a number of rogue waves embedded in an incoherent wave field.
\begin{figure}
\centerline{\includegraphics[width=9cm]{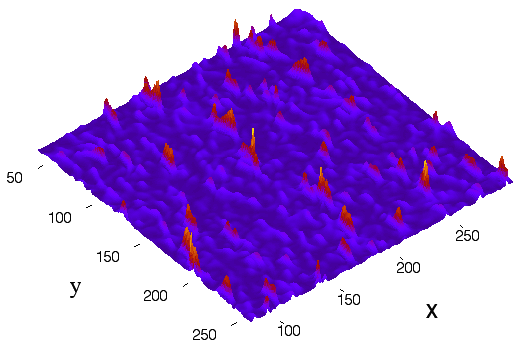}}
\caption{$|\psi|^2$  as a function of $x$ and $y$ at the time of maximum kurtosis,
 see Fig. \ref{fig_kurt_hyperbolic}.
 }
 \label{fig_rogue}
 \end{figure}
%

Before concluding, we find opportune to make a comment on the evolution of the spectral bandwidth.
So far, we have discussed an analytical result which provides an interesting perspective
on the generation of heavy tails.
However, equations (\ref{kurtosis})
and (\ref{kurtosis_2D})
are not closed: an evolution equation for the spectrum is still required. 
The present 
approach consists in considering the weakly nonlinear limit and derive the standard four-wave 
{\it wave kinetic equation} (see \cite{nazarenko2011wave}) from the (non-integrable) NLS type of equation using the wave turbulence theory.  
In such an equation the linear energy  is a constant of 
motion; this is a consequence of the fact that the transfer of energy and number of particle is ruled by exact resonance interactions. Therefore, even though the  spectrum may evolve, the spectral bandwidth  (related to the quadratic contribution to the Hamiltonian density) remains constant. Thus, if one is interested in studying changes in statistical properties of the 
wave field,  the need of considering non-resonant interactions in the kinetic equation is essential, see 
\cite{janssen03,annenkov2006role,annenkov2015modelling,suret2011wave} for details on the subject. 

In conclusion, we have presented an identity for a class of equations characterized by the NLS nonlinearity 
 that relates the  variation of the kurtosis with the variation of the spectral bandwidth. 
 In general, given an initial condition characterized by a narrow spectrum with random phases, the spectrum broadens due to the nonlinearity of the problem. 
 This broadening may be 
 accompanied by a change of the statistical properties of the wave field and, in particular, of the kurtosis. The latter weights the tails of the PDF.
 It should be noted that our approach is rather general as it can be applied whenever a conserved quantity of a partial differential equation contains a moment of the distribution. For example, in the  Korteweg-de Vries (KdV) equation, the Hamiltonian is directly connected to the skewness; therefore, given the evolution of the spectrum, a direct information on the asymmetry of the PDF of $u$ is available.
{\bf Acknowledgments} 
M.O. was supported MIUR Grant PRIN 2012BFNWZ2.
Dr. B. Giulinico is acknowledged for discussions.
P.S. and S.R.  were partially supported by the Labex CEMPI (ANR-11-LABX-0007-01).
The two-dimensional simulations were carried out on the High Performance Computing Cluster 
at the University of East Anglia.
\bibliography{references}
\end{document}